\documentclass[sigconf]{acmart}
\AtBeginDocument{%
  }

\copyrightyear{2026}
\acmYear{2026}
\setcopyright{cc}
\setcctype{by}
\acmConference[ASSETS '26]{The 28th International ACM SIGACCESS Conference on Computers and Accessibility}{October 25--28, 2026}{Vila Nova de Gaia, Portugal}
\acmBooktitle{The 28th International ACM SIGACCESS Conference on Computers and Accessibility (ASSETS '26), October 25--28, 2026, Vila Nova de Gaia, Portugal}
\acmDOI{10.1145/3797867.3829057}
\acmISBN{979-8-4007-2521-0/2026/10}

\begin{document}

\title{Towards Designing for (Dis)Trust in Technologies for Aging}


\author{Muhid Hassan Risvy} \orcid{0009-0002-8711-3165}
\affiliation{%
  \department{Department of Computer Science} \institution{New Jersey Institute of Technology}
  \city{Newark} \state{New Jersey}
  \country{USA}}
\email{mr2229@njit.edu} 

\author{Dimitri Theodoratos} \orcid{0009-0003-3510-8345} \affiliation{\institution{New Jersey Institute of Technology} \city{Newark} \state{New Jersey} \country{USA}} \email{dth@njit.edu} 

\author{Alisha Pradhan} \orcid{0000-0003-3419-9735} \affiliation{\department{Department of Informatics, Ying Wu College of Computing} \institution{New Jersey Institute of Technology} \city{Newark} \state{New Jersey} \country{USA}} \email{alisha.pradhan@njit.edu}







\renewcommand{\shortauthors}{Risvy et al.}

\begin{abstract}
Across cohorts of older adults and generations of technology, (dis)trust has continued to shape the acceptance of technologies for aging. Yet, we still know little about if or how to design for older adults’ distrust in technology. Towards this, we reviewed empirical studies in which older adults discussed trust and distrust in relation to technology. Our analysis shows that (dis)trust is constituted in three key ways: emerges when  material arrangements of digital systems become illegible, such as when older adults cannot determine where information is stored or trace its movement; is temporally constituted and fluid, building and shifting through prior experience and breakdowns, and at times distrust can operate as boundary work through which individuals limit technology's reach into their autonomy, relationships, and everyday lives. Our work yields design directions for supporting trust across different phases of technology use and by designing systems that make the movement and persistence of information more perceptible.
\end{abstract}



\maketitle

\section{Introduction}

Over the last several decades, researchers and designers have developed a plethora of technologies to support older adults' homes, health, or daily routines, ranging from smart home and assisted living systems (e.g., \cite{Caldeira2023a, Coughlin2007, Pal2018a}), to assistive robots (e.g., \cite{Zvanut2024c, Gul2024b}). These technologies hold considerable promise for supporting people as they age. Yet, older adults often remain hesitant to accept these technologies, for reasons including lack of perceived usefulness (e.g.,\cite{Selwyn2003,Lazar2016,Neven2010}),  stigma encoded in technologies for aging (e.g., \cite{Caldeira2022, Pradhan2020a}) and distrust toward technology (e.g., \cite{Knowles2018, Saka2025, Pena2021}). While research in human-computer interaction (HCI) and aging has productively responded to concerns around usefulness and stigma by foregrounding older adults’ agency through co-design and participatory approaches (e.g., \cite{Zhao2026, Zhao2025, Harrington2019}) or by empowering older individuals to build their own technologies (e.g., \cite{Ambe2019, Rogers2014}), designing for older adults' (dis)trust with respect to technology remains underexamined. 

Knowles and Hanson's work~\cite{Knowles2018}, among the few studies to interrogate what distrust means for older adults, found that distrust may function as a form of protest, signaling that important values such as fairness, social responsibility, and the burdens imposed by digitization on aging populations are at stake. Yet, when it comes to designing for distrust, many interventions currently frame it through broad concerns around data practices, privacy, and system reliability, thereby positioning distrust as a problem to be resolved through better transparency or privacy controls~\cite{Emami-Naeini2020, Nicholson2021, Wischnewski2023}. Despite these efforts, older adults’ distrust continues to surface even in systems explicitly designed around explainability and transparency (e.g., \cite{Mathur2026, Ehsan2021, Kim2023b}), suggesting how these interventions may not be enough. Importantly, what is striking is how distrust remains a persistent barrier to technology acceptance across cohorts of older adults and generations of technology: from early smart home research in the 2000s \cite{Coughlin2007} to contemporary studies of AI-based systems \cite{Saka2025, Chen2021, Desai2022} --- calling attention to a deeper design problem that we still perhaps do not fully understand. 

In this article, we take up this call by conducting a focused review of prior empirical studies in which older adults discussed trust and distrust in relation to technology. Our review of 59 empirical studies offers insights into how older adults’ distrust in technology is constituted: distrust emerges when material arrangements of digital systems become illegible, such as when older adults cannot determine where information is stored, trace its movement, or anticipate the outcomes of their actions within a system; distrust is temporally constituted and fluid, building and shifting through prior experience, breakdowns, and changing interpretations of technology; and distrust can operate as a form of boundary work through which older adults limit technology’s reach into their autonomy, relationships, and everyday lives. 
These findings contribute by extending current understanding of how older adults' distrust in technology is constituted. Specifically, our discussion yields novel directions for designing for distrust by: a) supporting trust across different phases of technology use rather than through discrete one-time interventions, b) designing for material legibility of information through systems that make the movement and
persistence of information more perceptible between both physical and
digital worlds, and c) understanding older adults' negotiable trust boundaries and designing around it.

\section{Method}

\begin{figure*}[h]
\centering
\includegraphics[width=0.62\textwidth, alt={A flowchart of the paper selection process. Searches of the ACM Digital Library and Google Scholar yielded 970 unique records, which were screened against two inclusion criteria requiring first-hand empirical accounts from older adults and trust-related terms in the findings, resulting in a final corpus of 59 empirical studies.}]{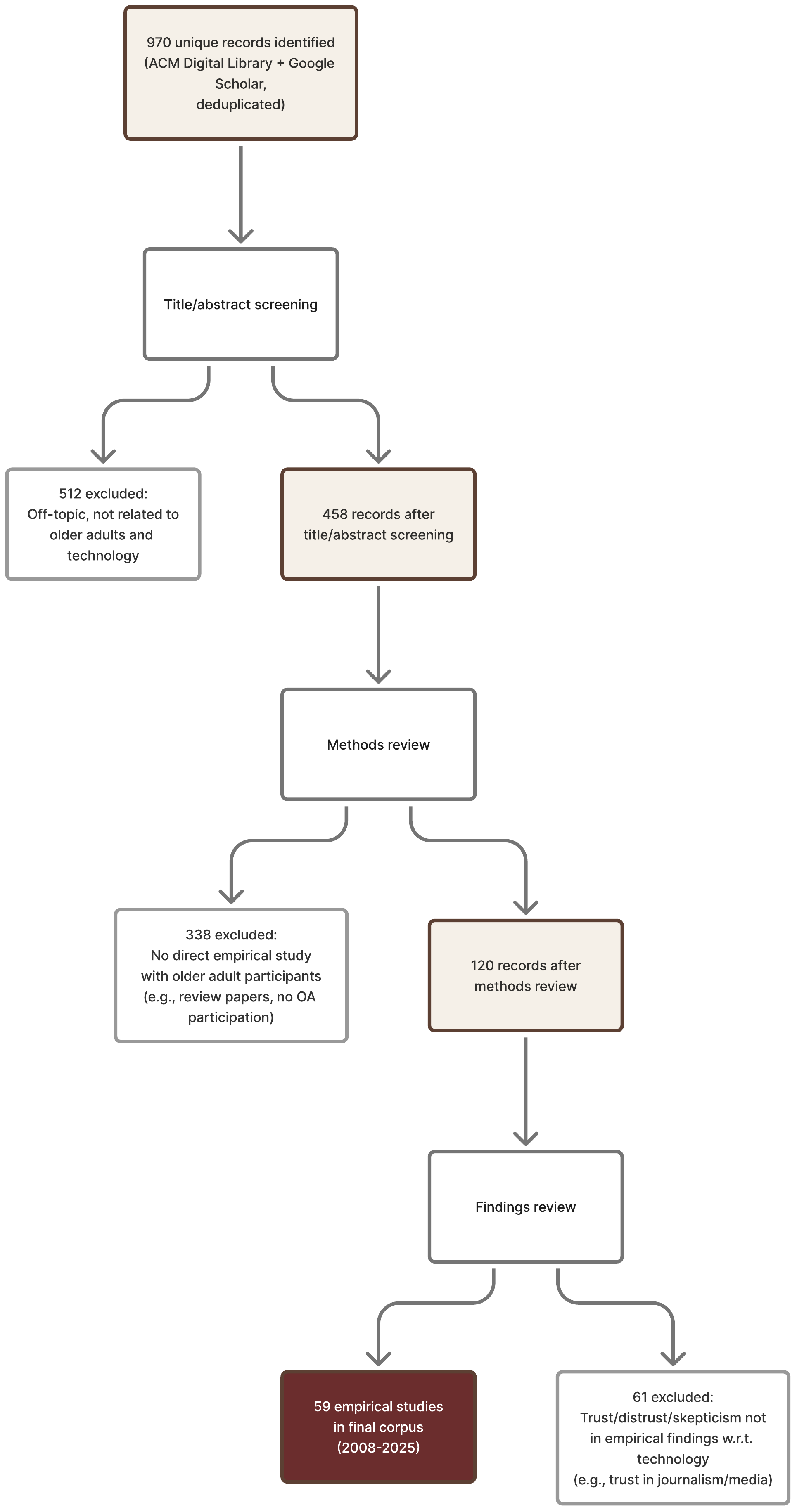}
\caption{Overview of the paper selection and analysis process, from 970 unique records to the final corpus of 59 empirical studies.}
\Description{A flowchart of the paper selection process. Searches of the ACM Digital Library and Google Scholar yielded 970 unique records, which were screened against two inclusion criteria requiring first-hand empirical accounts from older adults and trust-related terms in the findings, resulting in a final corpus of 59 empirical studies.}
\label{fig:selection-flowchart}
\end{figure*}

For data collection, we searched the ACM Digital Library and Google Scholar using population terms \textit{older adults}, \textit{seniors}, and \textit{elderly} with \textit{(trust OR distrust OR skeptic)} and \textit{technology}. While it is likely that adjacent terms from older adults' technology acceptance literature (e.g., refusal) may intersect with (dis)trust, broadening the search to include these terms risked shifting the data toward other factors shaping rejection, such as stigma  \cite{Neven2010} or perceived usefulness \cite{Selwyn2003, Lazar2016}. Moreover, distrust may appear in unexpected ways: not simply as rejection, but also as protest or a stance people hold while continuing to use technology~\cite{Knowles2018}. As such we scoped our search keywords around trust, distrust, and skepticism in line with our goal of understanding what older adults mean in their empirical accounts of (dis)trust and how to design for it. For ACM Digital Library, we directly searched using these terms; for Google Scholar, we used Publish or Perish (\cite{Harzing2010}, akin to prior work~\cite{PoP_CHI_example1, PoP_CHI_example2}), where each query returns up to 100 results sorted by a paper's citations. This resulted in 970 unique papers. Two researchers looked through a random subset of these papers' titles and abstracts and noted that not all papers provided first-hand empirical perspectives of trust from older adult participants in the context of technology. Discussions within our team led to two inclusion criteria for scoping the analysis to papers that a) reported first-hand empirical accounts from older adults through interviews, surveys, focus groups or similar methods; b) mentioned trust, distrust, or skepticism toward technology in findings of the paper. Therefore, we excluded review papers (e.g.,~\cite{Zhou2024}), papers discussing trust in journalism or media more broadly (e.g., \cite{Munyaka2022}), and papers in which trust-related terms did not appear in the empirical findings (e.g.,~\cite{Farina2019}). This  yielded a final corpus of 59 empirical studies published between 2008 and 2025. Data was collected in early February 2026. Table~\ref{tab:corpus} in Appendix has additional details about our corpus and Figure~\ref{fig:selection-flowchart} shows the paper selection process.

For each included paper, we extracted their findings passages reporting older adults’ trust-related attitudes, experiences, and included both participants' quotes and author-reported accounts (e.g., participants voiced, mentioned), which we qualitatively analyzed following a reflexive thematic analysis approach~\cite{Braun2019}. 
We followed an iterative process of coding the empirical data from these papers and theme development. Our codes included a mix of deductive codes identified from prior literature (e.g., loss of control over personal data) and inductive codes that emerged from data (e.g., materiality of data storage, fixed beliefs and instinctual distrust). See Table~\ref{tab:codebook} in Appendix for the full set of codes, and Table~\ref{tab:theme-codes} for how these codes connect to our three themes; read together, the two tables indicate the papers associated with each theme. Two researchers used this codebook to code a random subset of 20 papers while writing analytic memos. Given the interpretivist stance of our approach, measures such as inter-rater reliability were not calculated. Instead, the researchers discussed their analytic memos to identify primary themes in data. Through these discussions, codes were iteratively refined: for instance, initially separate codes such as ``fear of hacking'' and ``scam and unauthorized access'' were merged into a single code as we recognized they captured the same underlying concern about vulnerability to external breach. Example memos at this stage that speak to some of the final themes include ``trust related to material aspects of locating information'' and ``distrust entangled with resistance'' (see Table~\ref{tab:memos} in Appendix for sample memos and their corresponding codes). The first author continued coding and analyzing the data through analytic memos. Within our research group we iteratively discussed these codes and memos to identify the three high-level themes explaining older adults' trust and distrust with respect to technology --- described below. 


Adhering to calls for reflexivity in interpretivist qualitative research \cite{Collins2019, Erete2018, Harding2004}, we describe our positionality and how it shaped this study. First, our decision to pursue this research was informed by over eight years of working with older adults. Across this work, we repeatedly encountered participants expressing distrust through the language of privacy, making it difficult to disentangle distrust from broader concerns about data, control, and non-use, especially from a design perspective (i.e., if/how to design for older adults' trust towards technology). Second, our methodological approach was shaped by critiques in HCI and aging research against treating older adults as a homogeneous population, often characterized through deficit-based assumptions as uniformly reluctant, incapable, or disinterested in using technology \cite{Vines2015, Durick2013}. This led us to include only studies reporting first-hand empirical accounts from older adults and to exclude work that inferred their perspectives without direct participant data (e.g., systematic reviews). Our inductive approach to characterize trust is informed by prior work. While some work formally defines trust (e.g., as an expectation of a technology's reliability and helpfulness~\cite{Mcknight2011}), research on trust and aging complicates these definitions, noting trust may also operate as a protest expressing deeper moral objections~\cite{Knowles2018}. These informed our decision to let participants' accounts shape how trust is constituted. Finally, our analysis of the data was shaped by scholarship that treats older adults' rejection and distrust not as deficits to be overcome, but as meaningful stances grounded in their values, lived experience, and social worlds \cite{Neven2010,Lazar2016,Waycott2016,Caldeira2022}.

\section{Findings}

Across the studies we reviewed, older adults’ trust and distrust in technology were shaped less by isolated features than by broader ways of making sense of technology in everyday life. Three themes cut across the corpus: distrust emerging through how information systems became materially legible or illegible, trust as temporally constituted and fluid, and distrust as a form of boundary work through which older adults limited technology’s reach into autonomy, relationships, and daily life. 

\subsection{Material Legibility of Information Systems Producing (Dis)Trust} 

Across domains ranging from password management~\cite{Ray2021} to health applications~\cite{Desai2022, Pywell2020}, smart home devices~\cite{Saka2025}, and social media~\cite{Zhang2020}, studies with older adults often report distrust when individuals perceive a loss of control over personal data (noted in  N=11 papers in our corpus), fear hacking and unauthorized access (N=12), or feel uncertain about who receives their information (N=11). Read through Dourish’s materialist account of digital information \cite{Dourish2017}, however, these concerns point to a more specific phenomenon: distrust emerged when the material arrangements of information became difficult to account for. Here, materiality refers to the concrete properties of digital representations, such as their format, persistence, fragility, and transparency \cite{Dourish2017}. Rather than treating digital information as immaterial, Dourish argues that information is always encountered through particular forms of representation, transmission, and storage, and that these materialities shape how information can be understood and put to use \cite{Dourish2017}. In our corpus, this framing helps explain distrust not simply as a generalized response to privacy or security risk, but as emerging when the storage, movement, and actionability of information became materially illegible (i.e., when users could no longer trace where their data went, how it was stored, or what could be done with it)  in everyday life.


Across studies,  trust was closely associated with where information was stored, particularly with the material conditions of the storage platform. For example, in a study on older adults' password management practices, passwords written in a notebook were seen as ``safe'' with participants noting ``I always remember what a book is. And my book is safe''~\cite{Ray2021}. Because the information was contained in a known, physically present object, the participant could easily make sense of where the information was kept and how one would get to it, making the storage of information legible and easier to trust. 
Distrust towards systems that rendered storage of information less legible was seen in other studies as well. In Lehtinen's study, one participant described the cloud as ``a bit floating, like clouds are. You don't know where it's saved [...] I don't somehow trust that you could save things in it''~\cite{Lehtinen2023}. Here, cloud as an information storage platform could not be materially located or clearly imagined, further evidenced in older adults questioning the future existence of the cloud: ``how do you know they're gonna have the cloud tomorrow?''~\cite{Thangaraj2025}. Over decades, older adults have witnessed storage media become obsolete over their lifetimes, from floppy disks to USB drives \cite{Thangaraj2025}, and this lived experience of material impermanence of storage devices appeared to shape their distrust.  Across these accounts, distrust emerged when the material conditions of storage were no longer legible, that is, when people could not tell where information was kept (location) or whether that storage would endure (persistence). Notably, this distrust did not always produce outright non-use. Instead, it resulted in selective reliance, where participants used digital systems while withholding full dependence on them. For instance, older adults continued using a password manager, but explained, ``there can always be an error...That's why I keep a back-up... the fact that I have a back-up says I don't trust it''~\cite{Ray2021}. 

Distrust also emerged when the flow of information became materially illegible, that is, when participants could no longer tell where their information was going, who encountered it, and what happened to it once it moved beyond the device or participants' home~\cite{Desai2022, Wang2019, Mogavi2024, Saka2025}. For example, one participant in a study explained, ``I trust things less... information being passed on to other people and you don't really know who is looking at what anymore''~\cite{Knowles2018}. Here, distrust was tied to the sense that information no longer remained within a bounded interaction and the inability to trace its onward movement or identify who might encounter it. Similar concerns appeared with smart home devices, where a participant described Alexa as dangerous because ``all [my private speech] is recorded outside. People can listen to what's going on in your home illegally''~\cite{Chen2021}. Here distrust was not only associated with recording, but also with the material dislocation of speech: what begins inside the home is imagined as leaving that protected setting and entering an opaque external space, becoming ``unclear who's getting the data of what's said and what they're using it for''~\cite{Chen2021}. Across these accounts, distrust emerged at the point where the flow of information became untraceable.

A third, though less common, way in which material illegibility of information produced distrust was when outcomes of actions (e.g., clicking a link, scanning a QR code) could not be anticipated~\cite{Zou2024, Zhang2025}. Older adults distrusted email unsubscribe links because they could not predict what would happen upon clicking \cite{Zou2024};  in another study individuals distrusted QR codes because, ``you don't know what might happen after scanning QR codes. I'm afraid of being charged or that they might do something to the phone after scanning''~\cite{Zhang2025}.  In both these instances, distrust appeared to stem from interfaces that required an action whose consequences could not be previewed or inspected, especially when the destination of that action remained materially illegible.


\subsection{(Dis)Trust is Temporally Constituted}

Prior work notes that older adults' trust in technology is shaped by prior experiences with technology rather than at the point of first encounter~\cite{Kim2016, Zou2024, Ray2022, Willatt2024, Starkhammar2008}. This body of work suggests that trust is shaped by several temporal factors  such as duration of exposure~\cite{Tsai2020}, or repeated positive or negative interactions~\cite{Ray2022, Soubutts2025}. In some cases, these temporal histories extend to include institutional and racial histories of harm shaping how older adults interpret present systems~\cite{Willatt2024, Baseman2025, Seo2017, Harrington2023}. Contributing to these discourses, we highlight two salient temporal aspects shaping trust. First, across our corpus there was a notable asymmetry between how positive and negative experiences with technology shape trust. Second, trust is temporally fluid: it can strengthen, weaken, or shift as participants' circumstances change. 

Trust and distrust are shaped over time through prior experience, though not always symmetrically. In the studies we reviewed, positive experiences often supported trust through continuity with familiar brands or technology types, whereas negative experiences were more readily carried into later encounters with other technologies. When prior experience positively shaped trust, participants typically referred to experience with a similar technology or the same brand. For instance, Zou et al.\ found that prolonged use of antivirus software built trust: ``I've used Norton for so long... I have trusted them''~\cite{Zou2024}. In another study, participants who had used Microsoft products in professional settings were open to trusting other Microsoft products in personal contexts later in life~\cite{Ray2022}.
Across these positive cases, trust built gradually through continuity and familiarity, and it tended to remain within the boundaries of a recognizable brand or technology type.
By contrast, when prior experience shaped distrust, it could stem from negative experiences with a wide range of technologies. For example, one participant's Facebook account was hacked, and this experience contributed to distrusting  her smartphone: ``maybe I'm unnecessarily cautious, but I'll never even say a password out loud any more, because I know smartphones can listen in''~\cite{Soubutts2025}. As another example, a prior negative experience with workplace technology shaped reluctance to trust an everyday computing device such as an iPad~\cite{Willatt2024}.  These examples suggest that while trust often grew through continuity with familiar systems, distrust was more readily carried across technological contexts into later encounters with other devices.

Trust was also temporally fluid: it could strengthen, weaken, or shift as participants had new experiences, encountered breakdowns, or reinterpreted the role of technology in their lives. In some cases, technologies that were initially distrusted became more trusted as participants incorporated them into everyday routines. For example, several older adults in a study initially resisted mobile phones given to them by their children~\cite{McLean2011}. Yet after several months, some learned to use it for everyday tasks (e.g., keeping track of their  schedule) and eventually trusted smartphones. Here, trust did not emerge instantly, but was gradually shaped through sustained use and a changing relationship to the device.
In other cases, trust weakened over time.  Referring to a payment portal hacking incident, a participant noted, ``I don't use PayPal to the extent that I did before because it was a real big concern for me,'' describing an incident that reduced their willingness to rely on the platform~\cite{Saka2025}.  
As another example, older participants in~\cite{Panda2025} suggested that other older adults’ trust in smartphones and online banking rested partly on the fact that ``they haven't been scammed yet''~\cite{Panda2025},  treating trust as conditional and vulnerable to disruption in future. A few studies further suggest that trust after a breakdown may not follow a single trajectory. For example, after a GPS-enabled phone failed during a walk, a participant with cognitive impairment lost trust in the device and stopped using mobile phone entirely for a while. Only after his partner introduced a new phone and it proved reliable over time did his trust begin to repair; as he put it, ``since it had been proven to work'' \cite{Lindqvist2013}.  


\subsection{Distrust as Boundary Setting}

The third way in which distrust emerged in our corpus was as a form of boundary work: to set and maintain limits on how far technologies should extend into individuals' autonomy, relationships, and everyday lives. We interpreted distrust as boundary setting when participants invoked it to limit what technologies were permitted to do in their lives. In these accounts, distrust marked attempts to preserve autonomy, maintain valued relationship dynamics, or resist forms of mediation participants experienced as intrusive or misaligned with their values.

In several studies, distrust was less about how a technology technically worked (as discussed in 3.1 and 3.2) and more about setting boundaries by resisting what its use might authorize and how it might mediate their relationships, autonomy, and everyday lives \cite{McLean2011, Elavsky2024, Diehl2022, Pradhan2020, Trajkova2020, Blok2020}. 
For instance, some older adults perceived mobile phones gifted by family as instruments of surveillance, protesting, ``I don't want to have my daughter keep checking on me. She'll never leave me alone; she'll torture me!''~\cite{McLean2011}. Here, the phone was rejected because it could become a mechanism through which the daughter’s care could become persistent monitoring: threatening not only the individual’s autonomy but also reshaping her relationship with her daughter on surveillant terms. Distrust also emerged as boundary work when older adults perceived overreliance on technology as introducing unwanted dependence into everyday life and, in turn, threatening their autonomy \cite{Diehl2022, Pradhan2020, Trajkova2020}. For example, a participant in \cite{Diehl2022} did ``not trust technology,'' despite understanding ``that technology is important and that it helps many people,'' because she believed ``that she would feel somehow trapped if using a cell phone'' and  was currently ``happier and freer without technology''~\cite{Diehl2022}. Fearing that smart home devices can introduce dependence on technology, some participants in \cite{Trajkova2020} felt that for ``people who are basically independent...Alexa could be an unfortunate thing''. In these instances, participants invoked distrust in ways that helped justify non-use and maintain boundaries around their autonomy.

Boundary work also appeared in how older adults resisted technologies that conflicted with their values~\cite{Mogavi2024, Kim2016, Zhang2020, Blok2020, Sin2022, Baseman2025, Knowles2018}.  
For example, in Mogavi et al.'s study~\cite{Mogavi2024}, one participant rejected ``[foreign] exergames'' because they did not align with ``our [home] values'' and were seen as a means to ``infiltrate our culture,'' with others in the same study adding, ``the old ways we know are fine''~\cite{Mogavi2024}. Similarly, participants in other studies described, ``I am probably a little bit stubborn because I do not trust the new ways''~\cite{Kim2016}.  Perceiving Facebook as conflicting with their  value driven ``reluctan[ce] to have the whole community know about [them],'' some participants in~\cite{Zhang2020} avoided using the platform. Here, distrust reflects an act of drawing boundaries around personal values and familiar practices participants wanted to preserve. Across this theme, distrust expressed through boundary setting more often appeared to lead to outright refusal or substantial non-use of technology: with participants rejecting phones entirely in~\cite{McLean2011, Diehl2022}, or rejecting ``the new ways''~\cite{Kim2016}. 

\section{Limitations}
Our approach to data collection and analysis has tradeoffs. Inline with our goal of reading through older adults' empirical accounts to understand how to design for distrust, this work offers an interpretive account of how older adults’ distrust in technology is constituted across multiple studies. Yet, it is likely that the corpus we analyzed may not be exhaustive (e.g., ranking biases of databases searched). Akin to other works focused on design operational insights (e.g.,~\cite{Webber2023}), we reviewed papers from ACM DL and Google Scholar. While this approach is likely to yield papers that talk about trust in the context of technology, design, and aging, this approach also risks excluding papers from other fields with lesser visibility (due to Google Scholar algorithm). Moreover, the search results on platforms like Google Scholar may not be reproducible. Given the interpretivist epistemological stance of the work and these methodological tradeoffs, we exercise caution and note that findings from this work are not intended to produce generalizable results, and as such our work should not be treated as a systematic review of literature. Next, our analysis draws on findings reported in published papers and has tradeoffs. While it allows us to include older adults' perspectives across a large number of studies over decades, it is likely that several important details relevant to informing (dis)trust and technology design may not be presented (e.g., if it does not fit a neat story or is beyond the scope of the larger research goals~\cite{Lazar2021}). Moreover, since the empirical data was not collected by researchers who analyzed the data, we are cautious in overstating our claimed contributions and instead intend the design ideas we discuss to be preliminary forays into designing for (dis)trust. 

\section{Discussion and Conclusion}

In reviewing older adults' empirical accounts of trust and technology from prior scholarship, we extend current understandings of how distrust is constituted beyond broad explanations such as privacy concerns, security risks, and usability barriers (e.g.,~\cite{Vaportzis2017, Morrison2021, Frik2023}). Specifically, we identify three distinct but related dynamics: distrust emerging through the material illegibility of information systems, the temporal constitution and fluidity of trust, and as a form of boundary work to preserve autonomy, relationships, and everyday life. 
Second, our findings also contribute by extending an understanding of the relation between distrust and technology use/non-use. Corroborating \cite{Knowles2018}, we find distrust cannot reliably predict non-use among older adults. While Knowles et al. \cite{Knowles2018} find that older adults may continue using distrusted technologies because they are expedient or necessary, our findings suggest that there appears to be a link between how distrust is produced and the resulting use pattern:  distrust produced by opaque material practices may result in selective use by minimizing full reliance on technology, while distrust emerging as boundary work may result in non-use. Most salient, however, is the temporal fluidity of this relationship: as trust changed, so did use. Older adults might adopt a technology, later hedge their reliance on it,  or withdraw from it altogether after a breakdown, suggesting how shifts in (dis)trust also shift trajectories of use and non-use over time. 

These findings on how trust is temporally constituted and fluid yield new directions for design. First, current design interventions intended to build trust focus on discrete fixes such as explanations, onboarding prompts, or warnings  \cite{Emami-Naeini2020, Liao2020, Wischnewski2023}. Our work instead suggests that \textit{mechanisms for supporting trust among older adults may need to look different across phases of use}: while an explanation or data-policy disclosure may help build trust for initial adoption, it may be less effective after a breakdown, where peer-based learning~\cite{Nicholson2021, Xing2024}, gradual re-engagement~\cite{Baughan2023}, or fallback options may better support trust repair.   
What this entails for future work is learning from older adults what mechanisms for trust repair would work best at different stages of technology use (initial introduction, early experimentation, routine use, after failure, and after broader harms). 
Second, existing approaches to building trust typically focus on the current system alone (e.g., a privacy label describing a single IoT device's data practices~\cite{Emami-Naeini2020}, or an explanation for a particular algorithmic output~\cite{Liao2020}). However, our findings show that older users may arrive with trust histories, especially distrust histories, indicating that trust-building cannot focus on the current system alone. A smartphone, for example, may be distrusted not because of its present design, but because of an earlier workplace technology failure. Designing for distrust, then, may require \textit{attending to the specific pathways through which (dis)trust was formed}, since the same technology may call for different forms of trust support depending on whether it is encountered through continuity, breakdown, prior harm, or inherited values.

Intelligent assistive technologies for aging present an important context for extending our findings on unpacking the temporality of (dis)trust. Prior work often surfaces boundary setting around intelligent technologies that can monitor and adapt, such as smart care technologies and voice-first ambient assistants that sense daily routines~\cite{Elavsky2024}: for instance, older adults physically unplugged devices to ensure they were not being listened to~\cite{Cuadra2023a} or bounded who could see their health data, e.g., a doctor but not family~\cite{So2024}. However, many of these reported findings are at moments of early deployment. Less is known about how older adults appropriate adaptive assistive systems over longer periods, especially as these systems learn from daily activity, change their behavior, and become entangled with shifting care needs over years. Here technology domestication can provide future work a fruitful framework to unpack how assistive technologies are domesticated, i.e., how they are integrated into everyday routines and relationships over time~\cite{Alizadeh2026}. Additionally, leaning on this domestication perspective, future work can further unpack how \textit{(dis)trust itself is domesticated} through ongoing use, breakdown, repair, and changing relationships to care.

Understanding (dis)trust through the material legibility of information, beyond broad concerns of privacy or loss of control over data  (e.g.,~\cite{Ghorayeb2021, Ghadamighalandari2025, Saka2025, Pena2021}),  yields additional insights for design. Existing trust-supporting interventions in HCI and usable security often emphasize clear explanations of system behavior (e.g., privacy nutrition labels~\cite{Emami-Naeini2020}, network traffic visualization tools~\cite{Huang2020, Seymour2020})  to help users make informed judgments. Our findings suggest that such explanations address only part of the problem. Across the studies, older adults not only asked what data a system collected~\cite{Ghorayeb2021} or why it produced a given output~\cite{Hao2024}; they were additionally trying to understand where information resided, how it traveled and what happened to it over time. Designing for trust, then, may require moving beyond static, text-based explanations  (e.g., privacy policies, cookie consent dialogs, or data disclosure notices~\cite{Feng2021}) to additionally \textit{designing representations that make the movement and persistence of information more perceptible in everyday life}. One possible direction is to explore hybrid or physical forms that render otherwise invisible data processes more legible. For example, physical objects can make data flow  perceptible through  dashboards that visualize data transmission~\cite{Windl2025} or hardware shutters can signal sensor states~\cite{Ahmad2020}. Such interventions can help older users reason not just about what a system does, but about the material conditions through which their information is stored, transmitted, and retained.

Above, we discussed designing for (dis)trust 
which is inline with the larger HCI goal of calibrating trust \cite{Kahr2024, Meng2021, Baughan2023}. Yet, older adults' act of drawing boundaries through distrust was not about whether a system was reliable or understandable, but whether it should be permitted to mediate particular relationships, values, or forms of dependence.  This, in turn, pushes us to ask \textit{whether} we really need to design for distrust as a deficit to overcome. This, in turn, speaks to prior work that cautions about treating distrust merely as an obstacle to repair can obscure the legitimate moral and societal concerns older adults are expressing, and that distrust can itself function as critique by pointing to flawed technologies~\cite{Knowles2018}.  In joining Knowles et al. we agree that older adults' attitudes towards trust can serve as learning points to identify potential issues with technology. Yet attuning to how older adults actively drew boundaries around technology, together with the fluidity of trust, can provide a productive lens for designing technologies for aging. We see an opportunity in understanding older adults' \textit{negotiable boundaries of trust}, in particular how it evolves with time, context, and life situations. For an older person, a negotiable boundary of trusting a digital reminder could be \textit{maintaining a parallel physical backup} (to avoid overreliance on digital systems entirely). This points to a different design orientation for aging-in-place technologies, one that respects such boundaries rather than attempting to override them. Such boundaries are unlikely to be uniform across technologies. The conditions under which individuals are willing to trust an anthropomorphic smart home device (e.g., \cite{Seymour2021}) may differ from caregiver-configured smart home services that monitor daily activities (e.g., eating, bathing)~\cite{Belloum2021}. While such tensions appeared only sparingly in our corpus, future work should examine how older adults negotiate trust boundaries around systems that represent them, make decisions about them, or share their data across care networks. Going forward, learning from older adults which aspects of a system are acceptable, negotiable under certain conditions, or non-negotiable could therefore yield actionable design insights.

\begin{acks}
This work was supported in part by the National Science Foundation under award IIS-2433350. Opinions expressed do not necessarily represent official policy of the Federal government. We are also grateful to the anonymous reviewers for their valuable feedback.
\end{acks}


\bibliographystyle{ACM-Reference-Format}
\bibliography{references1}


\appendix
\section{Appendix}
\label{sec:appendix-codebook}
\label{sec:appendix-corpus}

\begin{table*}[h]
\centering
\caption{Overview of the 59 empirical studies in our corpus by technology type, application domain, and publication venue. Several studies span multiple categories.}
\label{tab:corpus}
\small
\renewcommand{\arraystretch}{1.15}
\begin{tabular}{p{8.5cm} r}
\toprule
\multicolumn{2}{l}{\textbf{Panel A: Technology Types}} \\
\midrule
\textbf{Technology Category} & \textbf{Papers} \\
\midrule
Mobile phone/app & 22 \\
Web-based platform & 17 \\
Voice assistant & 5 \\
Robot & 5 \\
Banking platform & 5 \\
Smart home/IoT & 4 \\
Wearable & 4 \\
Chatbot & 4 \\
AI/ML system & 4 \\
Automated vehicle & 3 \\
Cloud services/storage & 3 \\
Social media & 3 \\
Password manager & 2 \\
Other (e.g. health monitoring system, antivirus software, email) & 17 \\
\midrule
\multicolumn{2}{l}{\textbf{Panel B: Application Domains}} \\
\midrule
\textbf{Domain Category} & \textbf{Papers} \\
\midrule
Healthcare & 25 \\
Social communication & 15 \\
Information seeking & 11 \\
Financial services/shopping & 10 \\
Safety/emergency & 8 \\
Entertainment & 6 \\
Home management & 4 \\
Transportation & 3 \\
Cybersecurity & 2 \\
Other (e.g., photo storage, end-of-life data planning, life administration) & 12 \\
\midrule
\multicolumn{2}{l}{\textbf{Panel C: Publication Venues (42 unique venues)}} \\
\midrule
\textbf{Venue} & \textbf{Papers} \\
\midrule
CHI & 7 \\
PACM HCI & 4 \\
DIS & 4 \\
Frontiers in Psychology & 2 \\
IJERPH & 2 \\
JMIR & 2 \\
MobileHCI & 2 \\
TOCHI & 2 \\
Other: ACM ASSETS, USENIX Security, UbiComp, PETS, CUI and so on.& 34 \\
\bottomrule
\end{tabular}
\end{table*}

\begin{table*}[h]
\centering
\caption{Codebook: Code groups and codes applied to excerpts from 59 empirical studies. The ``Papers'' column shows how many studies (out of 59) were coded with that code. The ``Studies'' column lists all papers coded with that code. Three studies in our corpus focus on older adults with dementia or mild cognitive impairment: Lindqvist et al.~\cite{Lindqvist2013}, Starkhammar et al.~\cite{Starkhammar2008}, and Zhang et al.~\cite{Zhang2025}.}
\label{tab:codebook}
\small
\renewcommand{\arraystretch}{1.15}
\begin{tabular}{p{2.6cm} p{4.4cm} r p{8.2cm}}
\toprule
\textbf{Code Group} & \textbf{Code Name} & \textbf{Papers} & \textbf{Studies} \\
\midrule
1. Privacy and Control & Loss of control over personal data & 11 & \cite{Chen2021, Desai2022, Knowles2018, Mogavi2024, Ray2021, Saka2025, Seo2017, Wang2019, Zhang2020, Zhang2025, Zou2024} \\
 & Lack of transparency in data practices & 9 & \cite{Chen2021, Desai2022, Harrington2023, Knowles2018, Lehtinen2023, Mogavi2024, Pywell2020, Saka2025, Wang2019} \\
 & Fear of hacking and unauthorized access & 12 & \cite{Chen2021, Soubutts2025, Knowles2018, Oppert2023, Ray2021, Ray2022, Saka2025, Ware2017, Wilson2023, Wong2025, Zhang2025, Zou2024} \\
 & Surveillance and monitoring & 3 & \cite{Chen2021, Elavsky2024, McLean2011} \\
 & Distrust in data recipients & 11 & \cite{Chen2021, Desai2022, Knowles2018, Mogavi2024, Pywell2020, Saka2025, Seo2017, Wang2019, Zhang2020, Zhang2025, Zou2024} \\
\midrule
2. System Performance and Design & System unreliability & 15 & \cite{Asha2023b, Coughlin2007, Faber2020, Gul2024b, Lehtinen2023, Lindqvist2013, Nymberg2019, Oppert2023, Pradhan2020, Rahman2019a, Ray2021, She2026, Tsai2020, Tural2021, Willatt2024} \\
 & Deviation from expectations & 4 & \cite{Lindqvist2013, Nymberg2019, Sin2022, Zhu2025} \\
 & Deceptive design patterns & 5 & \cite{Aung2024, Harrington2023, Knowles2018, Morrison2021, Zou2024} \\
 & Transparency and explanation & 9 & \cite{Asha2023b, Aung2024, Soubutts2025, Kim2025, Knowles2018, Pywell2020, She2026, Sin2022, Wong2025} \\
 & Anthropomorphic and voice design & 4 & \cite{Brewer2023a, Cuadra2023a, Giorgi2025, Pywell2020} \\
 & Interaction modality and perceptual trust cues & 3 & \cite{Harrington2023, Pradhan2020, Pywell2020} \\
 & Platform instability & 2 & \cite{Ray2021, Thangaraj2025} \\
 & Failure to design for older adult needs & 5 & \cite{Elavsky2024, Nymberg2019, Seo2023a, So2024, Zhu2025} \\
 & Advertisements undermine trust & 3 & \cite{Knowles2018, Morrison2021, Sin2022} \\
 & Materiality of data storage & 4 & \cite{Chen2021, Lehtinen2023, Ray2021, Thangaraj2025} \\
\midrule
3. Experience, Familiarity, and Knowledge & Positive experiences build and sustain trust & 10 & \cite{Aung2024, Kim2016, Lindqvist2013, McLean2011, Panda2025, Rahman2019a, Ray2021, Ray2022, Starkhammar2008, Zou2024} \\
 & Negative experiences erode trust & 19 & \cite{Aung2024, Chen2021, Harrington2023, Knowles2018, Lindqvist2013, Nymberg2019, Morrison2021, Oppert2023, Ray2021, Saka2025, Seo2023a, So2024, Thangaraj2025, Tural2021, Ware2017, Willatt2024, Zhang2025, Zhu2025, Zou2024} \\
 & Knowledge gaps create distrust & 13 & \cite{Harrington2023, Kim2025, Knowles2018, Lehtinen2023, Nymberg2019, Oppert2023, Pywell2020, Ray2021, Ray2022, Seo2017, Tural2021, Zhang2020, Zhang2025} \\
 & Fixed beliefs and instinctual distrust & 17 & \cite{Asha2023b, Elavsky2024, Harrington2023, Kim2016, Knowles2018, Nymberg2019, Mogavi2024, Oppert2023, Pradhan2020, Ray2021, Ray2022, Seo2017, Starkhammar2008, Tural2021, Willatt2024, Zhu2025, Zou2024} \\
\midrule
4. The Social Ecosystem of Trust & Family as trust mediators & 3 & \cite{Blok2020, Lindqvist2013, Mendel2021} \\
 & Trusted sources (human, institution) enable adoption & 11 & \cite{Asha2023b, Cajita2018, Harrington2023, Knowles2018, Momeni2018, Pywell2020, Sin2022, Vaportzis2017, Wagner2021b, Wong2025, Zou2024} \\
 & Peer and community influence on trust & 7 & \cite{Baseman2025, Chen2021, Knowles2018, Mendel2021, Tural2021, Ware2017, Zou2024} \\
 & Organizational reputation and scale & 6 & \cite{Desai2022, Knowles2018, Pywell2020, Wagner2021b, Wong2025, Zou2024} \\
 & Institutional failures propagate distrust & 4 & \cite{Harrington2023, Knowles2018, Nymberg2019, Ware2017} \\
 & Others' negative experiences compound distrust & 5 & \cite{Chen2021, Knowles2018, Tural2021, Ware2017, Wong2025} \\
\midrule
5. Personal Identity and Values & Value conflict creates distrust & 6 & \cite{Kim2016, Kim2025, Knowles2018, Mogavi2024, Sin2022, Zhang2020} \\
 & Cultural norm misalignment & 4 & \cite{Brewer2023a, Gul2024b, Knowles2018, Mogavi2024} \\
 & Self-distrust & 9 & \cite{Blok2020, Harrington2023, Knowles2018, Nymberg2019, Pradhan2020, Seo2017, Seo2023a, Vaportzis2017, Ware2017} \\
 & Perceived usefulness builds trust & 12 & \cite{Chong2023, Ejdys2018, Elavsky2024, Lehtinen2023, Lindqvist2013, McLean2011, Oppert2023, She2026, Starkhammar2008, Vaportzis2017, Zhu2025, Zvanut2024c} \\
 & Future care needs increase trust willingness & 1 & \cite{Ejdys2018} \\
 & Fear of losing autonomy & 4 & \cite{Asha2023b, Diehl2022, Panda2025, Pradhan2020} \\
 & Physical risk amplifies distrust & 5 & \cite{Asha2023b, Coughlin2007, Faber2020, Momeni2018, Willatt2024} \\
 & Demographic factors & 5 & \cite{Baseman2025, Brewer2023a, Ejdys2018, Harrington2023, Rahman2019a} \\
\midrule
6. Consequences of Trust and Distrust & Avoidance of high-stakes tasks & 6 & \cite{Momeni2018, Pradhan2020, Saka2025, Seo2017, Ware2017, Wilson2023} \\
 & Distrust suppresses adoption broadly & 21 & \cite{Chen2021, Harrington2023, Kim2016, Knowles2018, Lindqvist2013, McLean2011, Oppert2023, Pywell2020, Ray2021, Saka2025, Seo2017, Seo2023a, Sin2022, Tsai2020, Wang2019, Ware2017, Willatt2024, Wong2025, Zhang2020, Zhang2025, Zhu2025} \\
 & Distrust as protest and strategic non-use & 2 & \cite{McLean2011, Mogavi2024} \\
 & Selective trust across services & 7 & \cite{Aung2024, Oppert2023, Giorgi2025, Ray2022, Wagner2021b, Zhang2020, Zou2024} \\
 & Anxiety-avoidance feedback loop & 5 & \cite{Nymberg2019, Oppert2023, Tsai2020, Wong2025, Zhang2025} \\
 & Not caring about data privacy & 1 & \cite{Zvanut2024c} \\
\bottomrule
\end{tabular}
\end{table*}

\begin{table*}[h]
\centering
\caption{Connection between the themes and the codes (Table~\ref{tab:codebook}). Reflecting the interpretivist stance of our analysis, the themes are generated reading multiple codes together and a code may connect to more than one theme, while some codes do not connect to any theme.}
\label{tab:theme-codes}
\small
\renewcommand{\arraystretch}{1.3}
\begin{tabular}{p{5.4cm} p{10.4cm}}
\toprule
\textbf{Theme} & \textbf{Code Names} \\
\midrule
1. Material Legibility of Information Systems Producing (Dis)Trust &
Loss of control over personal data \newline
Lack of transparency in data practices \newline
Fear of hacking and unauthorized access \newline
Distrust in data recipients \newline
Materiality of data storage \newline
Deceptive design patterns \\
\midrule
2. (Dis)Trust is Temporally Constituted &
Positive experiences build and sustain trust \newline
Negative experiences erode trust \newline
System unreliability \newline
Organizational reputation and scale \\
\midrule
3. Distrust as Boundary Setting &
Surveillance and monitoring \newline
Fear of losing autonomy \newline
Value conflict creates distrust \newline
Cultural norm misalignment \newline
Distrust as protest and strategic non-use \newline
Fixed beliefs and instinctual distrust \\
\bottomrule
\end{tabular}
\end{table*}

\begin{table*}[h]
\centering
\caption{Sample analytic memos and their corresponding codes from our coding process. Each row shows an excerpt from a paper in our corpus, an analytic memo written while coding it, and the resulting code(s) from the codebook (Table~\ref{tab:codebook}).}
\label{tab:memos}
\small
\renewcommand{\arraystretch}{1.3}
\begin{tabular}{p{1.6cm} p{5.7cm} p{5.3cm} p{3.2cm}}
\toprule
\textbf{Source} & \textbf{Excerpt from paper} & \textbf{Analytic memo} & \textbf{Code(s)} \\
\midrule
Lehtinen~\cite{Lehtinen2023} & ``Cloud services were sometimes used to store photos [\ldots] the participants seldom saw any value in the possibility to share the content or collaborate online and even questioned the reliability: `It's a bit floating, like clouds are. You don't know where it's saved [\ldots] I don't somehow trust that you could save things in it' (P12).'' & It's like because you can't see or feel the cloud, you can't trust it. This is something that comes up so much in research, but still we take storage on the cloud for granted (despite a big group not being able to trust it). & Materiality of data storage \\
\midrule
Willatt~\cite{Willatt2024} & ``[She] was initially reluctant to purchase clothes online [\ldots] This exchange about fabric prompted Pandora to tell a story about her time working in a cloth factory -- a period that left her with what she described as a strong distrust of technology [\ldots] `Now you know where my distrust of technology comes from.''' & A participant traces her distrust of technology to traumatic experiences in a cloth factory, where automated machinery malfunctioned and seriously injured workers, and caused the participant's own hearing loss. & Negative experiences erode trust \\
\midrule
McLean~\cite{McLean2011} & ``Several participants in our study felt distrustful of the mobile phones their children had bought them. They either abandoned them, or adopted them only gradually after becoming personally persuaded of their value. `I don't want to have my daughter keep checking on me,' one man frankly stated. `She'll never leave me alone; she'll torture me!''' & Older adults distrust mobile phones given by their children because they perceive them as monitoring devices. But once they see usefulness, they would see it as a less intrusive device. & Surveillance and monitoring \newline Perceived usefulness builds trust \\
\bottomrule
\end{tabular}
\end{table*}

\end{document}